\documentclass{article}
\usepackage{graphicx}
\usepackage[autostyle]{csquotes}
\title{Space Astronomy at TIFR:\\ From Balloons to Satellites\thanks{Based on a talk given on 2022 December 17 during the conference `Landmarks@75', organised by TIFR Alumni Association.
}}
\author{A. R. Rao  \\
{\it Email: arrao@tifr.res.in a.raghu.rao@gmail.com}\\
	Department of Astronomy and Astrophysics  \\
	Tata Institute of Fundamental Research, Mumbai 400005, India \\
		}

\date{}
\begin{document}

\maketitle

\begin{abstract}
Tata Institute of Fundamental Research (TIFR) has a very long tradition of conducting space astronomy experiments. Within a few years of the discovery of the first non-solar X-ray source in 1962, TIFR leveraged its expertise in balloon technology to make significant contributions to balloon-borne hard X-ray astronomy. This initial enthusiasm led to extremely divergent all-round efforts in space astronomy: balloon-borne X-ray and infrared experiments, rocket and satellite-based X-ray experiments and a host of other new initiatives. In the early eighties, however, TIFR could not keep up with the torrent of results coming from the highly sophisticated satellite experiments from around the world but kept the flag flying by continuing research in a few low-key experiments. These efforts culminated in the landmark project, AstroSat, the first multi-wavelength observatory from India, with TIFR playing a pivotal role in it. In this article, I will present a highly personalised and anecdotal sketch of these exciting developments.
\end{abstract}
~
\section{Space astronomy: probing the cosmos using the invisible rays}
 ~\\
 
 Space astronomy is the child of the space era.\\
 
Traditionally, the optical region is the only wavelength through which astronomical observations were possible. The tremendous development and insight obtained in Physics during the early twentieth century were duly applied to understand the vast amount of data obtained from stars using large optical
telescopes. Most of our concepts on how stars formed, evolved and radiated their energy came from this meticulous use of Modern Physics to optical spectral data. The understanding was so mature that the doyen of twentieth-century astrophysics Arthur Eddington claimed that \enquote{a civilisation completely covered with dust but knows laws of physics, can calculate the size, mass and temperature of stars}.

Such Archimedean intellectual arrogance (\enquote{Give me a lever long enough and a fulcrum on which to place it, and I will move the world}) perhaps is the result of the realisation that every single law of Physics discovered in this tiny planet called Earth, is applicable anywhere else in the vast cosmos, almost without exception. This last statement is simultaneously a profound as well as a trivial statement. Profound in the sense that the laws discovered by us tiny humans are applicable all through the universe. Trivial in the sense that we humans will not comprehend anything beyond what can be understood by the laws we created!

Of course, Nature always has the last laugh. The advent of Space Astronomy provided observations in several invisible wavelengths  and demonstrated that there are vast new features in the universe requiring new insights and understandings.

The full range of the electromagnetic radiation (shown in frequency at the bottom X-axis and as the energy of photons in the top X-axis) is shown in Fig 1. The plot shaded with graded colour shows the height (given in km in the right-hand scale and as a fraction of the atmosphere in the left-hand scale) that we have to go up to detect at least 50\% of the radiation incident at the top of the atmosphere. We have to use balloons, rockets or satellites to see radiations outside the visible band (shown as a shaded yellow region in the figure), mainly to observe infrared, X-ray, and gamma-ray photons.

The Planck black body radiation curves are shown for 6000 K and 1 million K temperatures in Fig 1. The former is close to the surface temperature of the Sun, demonstrating the utility of optical wavelength to observe Sun-like stars. A million-degree object, however, has its peak emission in X- rays. It was thought earlier that not many objects in the universe have enough energy to sustain emissions at such a high temperature. The observations during the space era have indeed shown the existence of strange objects like neutron stars and black holes, copiously emitting X-rays and gamma-rays. The enigmatic objects called gamma-ray bursts signalling the extreme states of compact objects, the birth of black holes during supernovae or death due to mergers accompanied by gravitational waves, emit the equivalent of the rest mass energy of the Sun in a matter of seconds, mainly in X-ray and gamma-rays.

\begin{figure}[h]
\centering
\includegraphics[angle=-90, width=\textwidth]{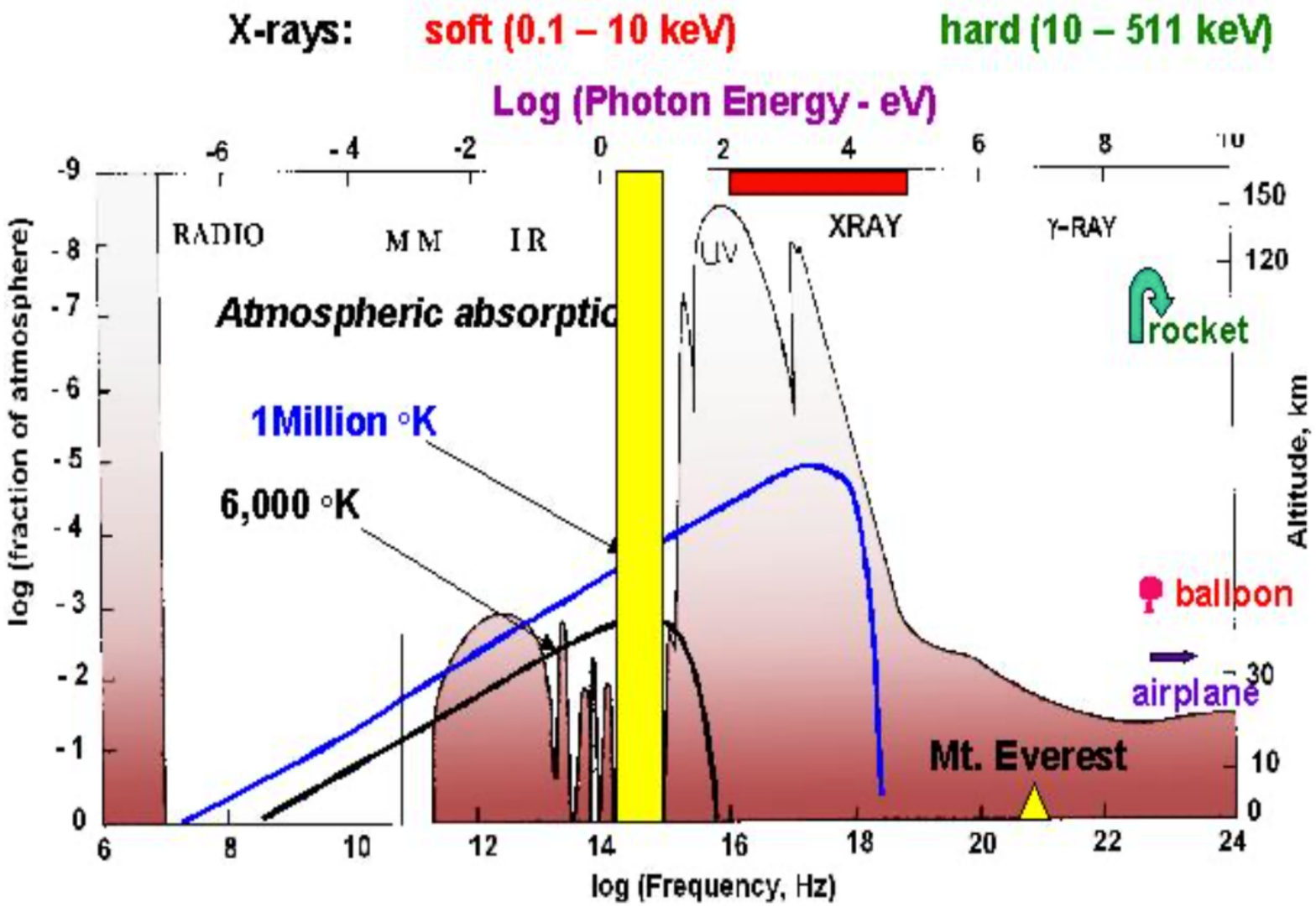} 
\caption{The full range of electromagnetic radiation is used for astronomy. The height above the Earth's surface at which the incident radiation at the top of the atmosphere is absorbed is plotted. It can be seen that the optical window is observable from Earth, and most other parts of radiation (infrared, UV, X-rays and gamma-rays) need observations at high altitudes, requiring the instruments to be lifted by balloons, rockets and satellites (Figure courtesy: K. P. Singh).}
\label{fig1}
\end{figure}

\section{The birth of Space Astronomy in the Apollo Era}

The seeds of space astronomy were sown post the second World War when the USA provided the captured German V2 rockets for scientific use. Rudimentary X-ray detectors like Geiger counters were used to measure X-rays from our nearest star, Sun. The amount of detected X-rays from the Sun, from the solar corona, is negligibly small compared to the optical emission. Hence, there was little expectation of detecting bright X-ray sources from other objects in the sky. The dream-like success of the space-faring efforts of the Soviet Union, starting with the launch of the first artificial satellite in 1957 and culminating in the spectacular and glorious moment of humans in space in 1961, resulted in the great space race of the sixties and the USA starting the famed Apollo program. The vast support provided to space-faring activities was used for scientific purposes. Riccardo Giacconi, the receiver of the 2002 Nobel Prize in Physics \enquote{for pioneering contributions to astrophysics, which have led to the discovery of cosmic X-ray sources}, was the one who revolutionised space astronomy with his incredibly long-term and visionary views on astronomy with the invisible rays accessible through space.

A trained Cosmic Ray physicist, Giacconi sensed the vast opportunities provided by the new initiatives in space technology. He was also highly influenced by the intellectual ambience provided by the noted Cosmic-ray physicist Bruno Rossi at MIT, and Giacconi headed the MIT spin-off company American Science \& Engineering (AS\&E). Collecting a band of young scientists, Giacconi embarked on a path of momentous discoveries. He improved the X-ray detection techniques. Though the then astrophysical understanding did not expect to find any non-solar X-ray sources in the sky bright enough to be detected by the modest X-ray detectors of that era, he boldly proposed to NASA to make a rocket flight to detect X-ray sources in the sky. It was turned own. He then re-proposed the same experiment to detect X-rays from the moon, which was approved, perhaps due to the core interest of NASA to send humans to the moon. During the now legendary rocket experiment of 1962,  Giacconi did not detect any X-rays from the moon but detected an extremely bright source called Sco X-1, thus heralding a fascinating journey of discovery in X-rays.

B. V. Sreekantan was at MIT during the sixties and was fortunate to witness the fascinating story of the birth and growth of X-ray astronomy. At that time, Homi Bhabha had collected a good number of young and enthusiastic scientists, and the mood and ambience at TIFR were one of attacking any scientific problems with a sense of adventure. TIFR had a very vigorous experimental activity of studying cosmic rays using balloon-borne instruments. Sreekantan was instrumental in using this infrastructure to start a balloon-borne X-ray program at TIFR. I joined TIFR in 1977, witnessed the growth of space astronomy at TIFR and was an integral part of it for more than four decades.  I will give below a personalised and anecdotal sketch of it. Subjectively, I have divided Space Astronomy at TIFR into three eras, Dreamers, Strugglers, and an Era of Renaissance.

\section{The era of Dreamers: 1967 - 1980}
   
  In the late seventies, there was a vibe of excitement in the research activities of Space Astronomy at TIFR. We young students were told the stories of the stalwarts in this field at TIFR.
  
\begin{figure}[h]
\centering
\includegraphics[angle=-90, width=\textwidth]{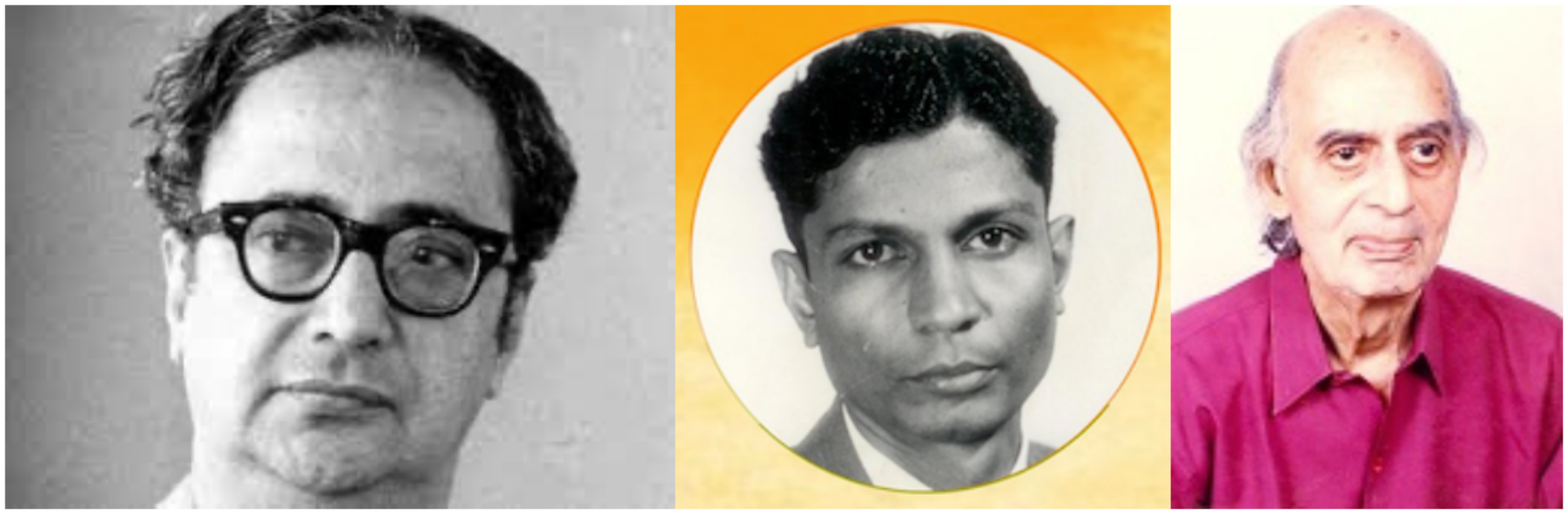} 
\caption{B. V. Sreekantan, R. R. Daniel and S. Biswas: the stalwarts of Space Astronomy at TIFR.}
\label{fig2}
\end{figure}

B V Sreekantan undoubtedly is the father of space astronomy at TIFR. He played an essential role in motivating and training many students. In the initial years, he proactively participated in balloon-borne X-ray astronomy experiments to get exciting results like the discovery of \enquote{Sudden Changes in the Intensity of High Energy X-Rays from Sco X-1} (Agrawal et al., 1969; Nature, 224, 51) and the detection of \enquote{Rapid Variations in the High Energy X-ray Flux from Cyg X-1} (Agrawal et al., 1971; Nature, 232, 38). After he became the Director of TIFR in 1975, he was the perfect benevolent fatherly figure to nurture and support all the new initiatives. He also navigated and managed the socio-political ambience to provide adequate support for all these new activities.

R R Daniel, the experimenter par excellence in cosmic rays, was also caught on the new initiatives of space astronomy. He was responsible for nucleating a group to start the difficult area of the balloon-borne infrared (IR) astronomy program at TIFR. The earlier statement by Eddington about the civilisation covered by dust deducing the nature of stars from basic Physics can be inverted to ask whether an open civilisation, like us, can deduce things about a problem enclosed in dust. Actually, not: the problem of the formation of stars from cool dust is opaque to us because we cannot make any measurements about it. IR region, sensitive to cold temperatures and the ability to penetrate dust, indeed was needed to fathom the intricacies of star formation. TIFR has made significant contributions to the problem of deciphering star formation through its balloon-borne activities initiated by R R Daniel.

S Biswas had the distinction of being the only Indian scientist whose innovative idea had an acceptance into the Space Shuttle program of NASA. He was an expert in using nuclear emulsions to measure the property of cosmic rays. The emulsions exposed to cosmic rays (emulsions taken to high altitudes by, say, balloons) record the particle tracks. The emulsions were later chemically treated to make the tracks measurable to provide the energy and direction of cosmic rays hitting these emulsions. Biswas had this brilliant idea of sending a stack of emulsions to space using the space shuttle and measuring these properties in the ground after their return. Additionally, the direction and energy of the particles can be used to infer their mass and charge if we have one further input. The experiment's cleverness uses a stack of emulsions with a very slow mutual rotation. One can get the incidence time by matching the cosmic ray tracks in different emulsions. This enables using the Earth's magnetic field configuration to figure out which mass and charge ranges are allowed such that these cosmic rays come from outer space navigating the Earth's magnetic field.

When I joined TIFR in 1977, we, the students, heard the stories of these stalwarts and got immersed in the new activities. Multiple  activities were going on. The Mark I version of the IR telescope, with a 30 cm mirror and six arc-minute pointing accuracy, already had several balloon observations carried out. The Mark IIa telescope, with a 75 cm mirror and a vastly improved pointing of 0.5 arc-minute, was getting ready for a balloon launch. Aryabhata, the first indigenous Indian satellite launched in 1975, had a component built by TIFR. Bhaskara satellite, to be launched in 1979, had an X-ray pin-hole camera built at TIFR. I had the exciting task of carrying a proto-type of the X-ray detector to the airport to hand it over to U R Rao, who, in turn, will take it to a NASA laboratory for further testing by a TIFR scientist there. As soon as I joined TIFR, I was dispatched to Thumba Rocket Station to test an X-ray detector for a soft X-ray rocket experiment. I was also deeply involved in designing and developing the largest area balloon-borne hard X-ray telescope.

The academic ambience was quite exciting. Though there was no formal graduate course structure, we were taught by stalwarts like J V Narlikar. An International Workshop on X-ray astronomy at TIFR had speakers like the renowned oratory teacher Walter Lewin. Listening to the scientific discussion between these stalwarts with our leaders like Krishna Apparao was in itself highly educative.

At this juncture, I would like to highlight the contributions of two individuals: P C Agrawal and S N Tandon.

Many, many people contributed during this era. I am highlighting these two individuals for two reasons: firstly, they, in their own way, influenced my research activities. P C Agrawal was my thesis guide and mentor. Shyam Tandon was a friend, philosopher, and well-wisher. Secondly, these two are the key personnel in the landmark TIFR activity of the AstroSat project: without the profound contributions of these two individuals, AstroSat would not have been possible.

\begin{figure}[h]
\centering
\includegraphics[angle=-90, width=\textwidth]{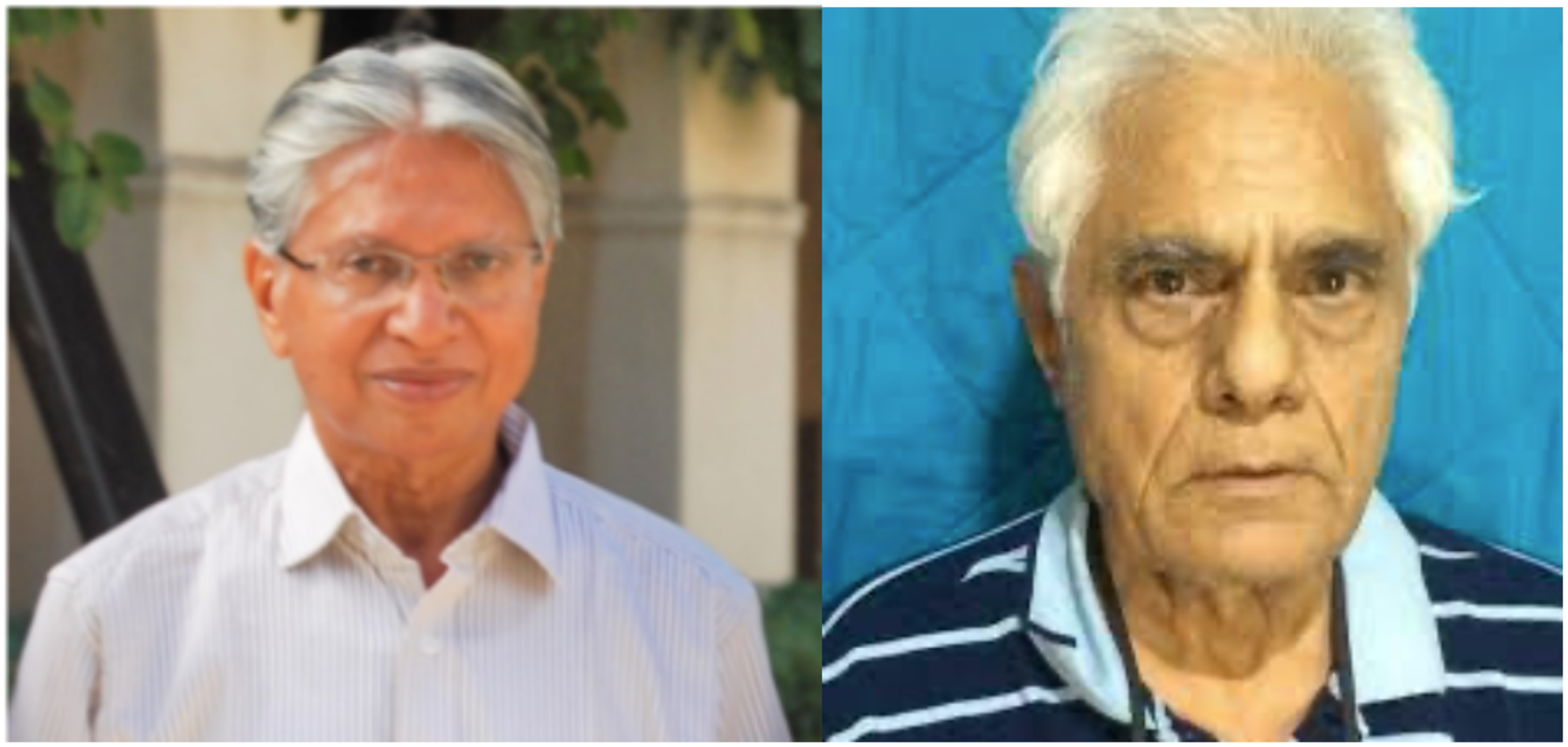} 
\caption{P. C. Agrawal and S. N. Tandon: the driving forces behind AstroSat.}
\label{fig3}
\end{figure}

After participating in the initial heady days of developing balloon-borne hard X-ray instruments (based on crystal scintillating detectors) for studying cosmic X-ray sources, P C Agrawal got trained to develop proportional counters at Caltech. He worked on the low energy experiment (A2) of the HEAO-A satellite of NASA. Armed with this valuable experience, he started an array of new space astronomy experiments based on proportional counters at TIFR. He developed a rocket experiment with K P Singh to study soft diffuse X-ray background; he started with R K Manchanda a large area Xenon filled proportional counters for balloon hard X-ray experiments (I joined him in this project as a graduate student); he made the X-ray pinhole camera for the Bhaskara satellite. His aggressive never-say-die spirit was highly motivating and contagious. When confronted with any serious problem, his instinctive approach is to suggest an aggressive solution and bang the table and say `why not' to any objections. He also had a deep sense of what was right and what could be done and could navigate the intricate political landscape of Indian science. The last two qualities proved extremely useful to realise AstroSat.

Shyam Narayan Tandon is an astute experimenter.

In the early days, X-ray astronomy at TIFR was done in an adventurous `can-be-done' way, and perhaps some sort of amateurism in the approach (everything had to be built in-house) was inevitable. These were ironed out later while building the AstroSat instruments due to our interaction with ISRO (In- dian Space Research Organisation). ISRO had learnt the importance of quality control the hard way (rockets will not go up without an extreme quality consciousness). Shyam was the one to realise this and imbibe this in the IR balloon payload early in his career. His meticulous approach to quality experiments was the prime reason for building the sophisticated IR telescope, highlighted by R Ramachandran (one of the TIFR alumnus, fondly called Bhaji) in the magazine Frontline in the eighties with a front page cover picture of the IR telescope. Additionally, Shyam's ability to approach any problem from first principles and arrive at a correct decision was truly inspiring and motivating to many of his associates.

\section{The era of Strugglers: 1980-2000}

Nevertheless, I will mark 1980 as the dividing line between the Dreamers era and the Strugglers Era.

There was some interesting work done at TIFR in the eighties too. Anuradha experiment was flown in Space Shuttle Spacelab-3 in 1985. The Mark IIb version of the balloon-borne IR experiment with a 100 cm mirror was built and operational during this period.

However, during this period, the world was galloping very fast, and it was too difficult for TIFR to catch up.

In the late seventies, NASA launched the high energy astronomical observatories (HEAO). HEAO-A, with an array of X-ray detectors, was launched in 1977. The extremely difficult task of focussing X-rays using grazing incidence telescopes was mastered, and it was used in HEAO-B (Einstein Observatory) and launched in 1978, improving X-ray detection sensitivities by several orders of magnitudes. The IR satellite of NASA, IRAS, was launched in 1983.

Sitting at the TIFR library facing the Arabian sea, we could read the special issues of the Astrophysical Journal Letters editions dedicated to the prime results from some specific instruments from many such satellites. New results and new insights were coming at regular intervals, and compared to these, our feeble attempts using balloon-borne instruments paled into insignificance. Indian instruments, satellites, and rockets were simply not mature enough to compete with such sophisticated results.

As one visiting astronomer commented at that time, `we were working in the backwaters of astronomy'.

   I vividly remember when K P Singh and I started jointly reading a collection of articles in a NASA memorandum called `X-ray astronomy in the 80s': a thick book containing review articles on various subjects opened up by the new X-ray observations. Every night, after dinner, we used to do a joint study for several days. We were used to studying X-rays coming from bright X-ray binaries: neutron stars or black holes accreting matter from a companion star. The superior sensitivity of Einstein Observatory opened up X-ray studies of a whole new set of objects: supernova remnants, white dwarf binaries, galaxies, clusters of galaxies, Active Galactic Nuclei, and blazars. X-ray astronomy had come of age. Reading all this exciting stuff highlighted the limitations of the results of our observations. And, something seemed actually to hit us. Each of us resolved to do something. Much later, K P Singh told me that it was during that time he had resolved to bring the X-ray focussing techniques to India.

\subsection{Multiple diverse activities}

In this era, however, several diverse activities indeed were carried out at TIFR. The balloon-borne IR telescope continued to make observations with better focal plane instruments, some with foreign collaborations. This group also started making inroads in ground-based optical observations. A program to use optical CCDs at various Indian optical telescope facilities was also initiated. R K Manchanda started building multiple X-ray detectors for balloon platforms. He provided ground support for the Italian Agile satellite, which provided some breakthrough observations on gamma-ray bursts.

S V Damle realised that we need to hitch up with space powers to make any meaningful contribution to space astronomy. He aggressively pursued col- laboration with the Soviet Union in a program called Gamma-ray Indo-Soviet Program (GRISP).

\begin{figure}[h]
\centering
\includegraphics[angle=-90, width=\textwidth]{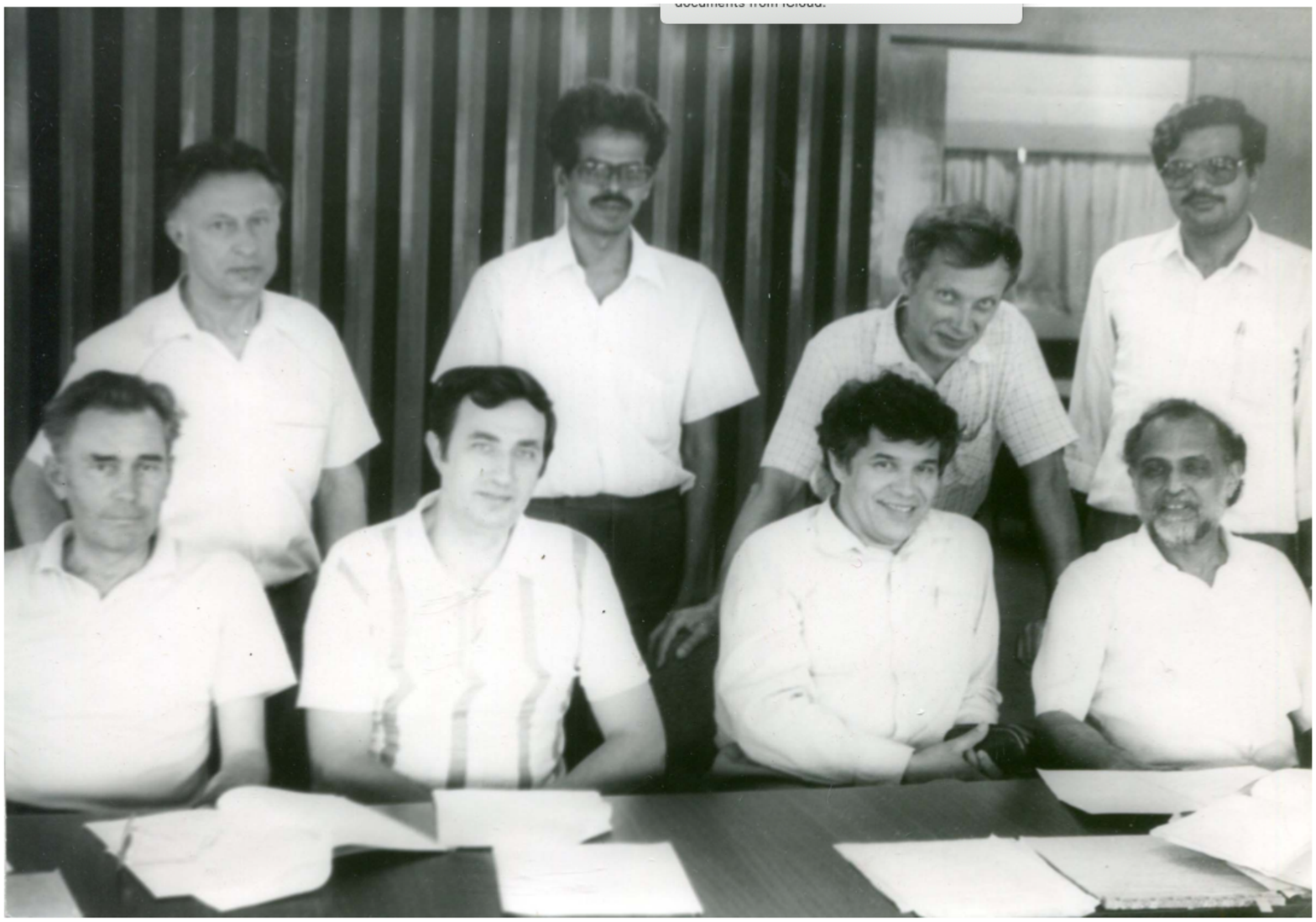} 
\caption{S V Damle (bottom right) with his colleagues from Gamma-ray Indo- Soviet Program (GRISP). Second from left (standing) is the author of this article.}
\label{fig4}
\end{figure}

The essence of building instruments for space astronomy is to get data and understand the working of the cosmos. There is a method to get this done without building instruments: get data from the best observatories worldwide based on proposals or directly get the archival data. Though this method makes the group incomplete and leaves a gap in the robustness of understanding the data, it can be used to strengthen the capacity to understand data from our own instruments. At that time, K P Singh had vast experience in analysing data from multiple space observatories and K P Singh, M N Vahia, and I started a project of meticulously analysing data from the European satellite EXOSAT. Within a year, we built a good reputation. K P Singh wanted this group work to extend to experimental activities too. We started an ambitious project to develop an X-ray focussing technique (the dream project of K P Singh) at TIFR. Unfortunately, this project was not funded due to the ostensible reason that even if such a telescope is built, there are no launch opportunities available in India.

\subsection{The Indian X-ray astronomy experiment, IXAE}

The eternally optimistic P C Agrawal continued with the hard X-ray balloon experiment using large-area proportional counters. He was consistently and continuously persistent with ISRO to provide launch opportunities. The nineties saw the development of the workhorse rocket of ISRO: the Polar Satellite Launch Vehicle (PSLV). The initial three rocket launches were development flights to demonstrate the rocket's reliability. In the third developmental flight, PSLV D-3, ISRO offered an opportunity to launch an X-ray payload. It was called the Indian X-ray astronomy Experiment (IXAE), and it was launched in 1996.

The decision to include this experiment in the satellite came quite late: there was much humming and hawing about whether TIFR was really ready to build an instrument in less than a year. Though we had a long experience building proportional counters, making a highly reliable and space-worthy instrument is a different ball game: we learnt it the hard way. Three pointed proportional counters (PPC) were designed and built at TIFR, and one X-ray Sky Monitor (XSM) based on the pin-hole camera principle was built by a close collaboration between ISRO and TIFR. It was highly satisfying that all instruments could be fabricated, tested, calibrated, and delivered within the rigid launch schedule.

The X-ray detectors worked. However, the spectral capability of PPC had some flaws: we could use the data as a two-band photometer without any spec- tral information. The XSM worked well, but the onboard data collection could collect only fifteen minutes of data per orbit, thus reducing the sensitivity drastically. Further, just a few months before the launch of IXAE, NASA had launched the Rossi X-ray Timing Explorer (RXTE). The PCA instrument of RXTE was similar in the design of PPC but had a five times larger area and complete spectral information. The ASM instrument of RXTE had a similar sensitivity as the XSM of IXAE, but there were three ASM units, and data were available all the time, unlike the fifteen minutes data from XSM. Further, IXAE was in a satellite in a polar orbit meant for Earth observations. This had two adverse effects. The X-ray background is very high near the polar regions (hence we can have data only about fifty per cent of the time). The satellite was devoted to Earth observations, and the sky observations required for IXAE were available only for a few months in a year.
  
  The situation looked hopeless.
  
Space astronomy is expensive, and each photon we collect in space is valuable. This was the era when the full capability of the internet was being exploited, and the data from RXTE was freely available. There was a vast expansion in the science that could be done with this type of X-ray timing instrument. Our know-how of X-ray data analysis from multiple international observatories was reinforced by our deep understanding of the instrument behaviour (`we could feel the photons hitting our instruments') and provided interesting results. We could use RXTE data to guide us on what sources to observe and use our understanding to extract valuable science. Even with the two-band photometry of PPC, the spectral measurements were used to infer the possibility of detecting matter advecting to a black hole. This result found a mention even in The Sunday Times.

 \section{The Era of Renaissance: 2000 onwards}
 
 The post IXAE ambience in the space astronomy group at TIFR was very vibrant. Using our own data in conjunction with the best available data from the world was very motivating to new students: a large number of students (TIFR students as well as from the university sector) participated in this activity. There were periodic workshops conducted at TIFR to attract and motivate students from a wider community. Several students from this time are faculties in other national Institutes and are currently contributing significantly to AstroSat.
 
The exciting scientific activity resulting from IXAE data impressed K Kasturirangan, the then ISRO chairman. He was a collaborator in many of the results from IXAE, and as an X-ray astronomer, he was fascinated by the new discoveries. He was also very impressed by the multifaceted abilities of our group at TIFR: we had the ability to design, develop, and deliver scientific instruments and also had the scientific mettle to analyse, interpret, and model the data: a sort of A-to-Z scientists.

P C Agrawal was pushing for a full-fledged astronomy satellite from India right from the launch of IXAE. Finally, with great help from Kasturirangan, it was approved in 2000 with financial support for initial instrument development.
 
 \subsection{AstroSat: the genesis}
   
   I cannot help repeating myself: AstroSat would not have been possible without the single-minded devoted work from P C Agrawal and S N Tandon.
   
Though the concept of AstroSat had complete backing from K Kasturirangan, P C Agrawal instinctively realised that bringing a full-fledged astronomy satellite into fruition required navigation through many unchartered territories. He was convinced that making a large area X-ray detector with superior sensitivity at higher energies is scientifically very rewarding, but, at the same time, it is also true that the science base, that is, scientists who can use such data, is relatively low in India. Getting a full-fledged astronomy satellite approved for a somewhat restricted science topic of measuring X-ray time variability from some select class of objects would have been challenging. India traditionally had a vibrant ground-based optical astronomy program, and it made good sense to include a 1-meter class optical/ UV telescope. However, a large area X-ray detector and an optical/UV telescope still lacked the punch. These instruments are based on vintage technologies: the know-how of making proportional counters and telescopes has been around for several decades. P C Agrawal co-opted K P Singh to make an X-ray focussing optics-based soft X-ray telescope (SXT): KP's dream project. New generation X-ray detectors like the near room temperature solid state device Cadmium Zinc Telluride (CZT) were taking the fancy of the scientific community. Sensing my interest in such detectors, P C Agrawal threw me to the deep end of the pool: he asked me to make a large area hard X-ray telescope using CZT detectors. Since ISRO was part of making the XSM for IXAE, an ISRO centre was co-opted to make an X-ray all-sky monitor.

Having a good configuration is only half the battle won. The technical complexities of the instruments, though within the grasp of the then-available Indian expertise, are too vast and numerous to be mastered and operationalised in a reasonable time. On the other hand, ISRO, over the years, has developed deep expertise in many engineering areas like mechanical design and analysis, thermal design, digital electronics, onboard software design, and precision fabrication with high quality and reliability. Properly harnessed, this expertise would be a boon for developing scientific equipment for space astronomy. For TIFR and ISRO to come together, there must be a method to merge and couple the contrasting styles of these eminent establishments. TIFR thrives on the individual brilliance of its scientists: an excellent boon for discoveries but, per- haps, a hindrance to developing a collective group culture to develop intricate space equipment. ISRO, on the other hand, is highly structured and hierarchal, a necessity for delivering rockets and satellites in scheduled time scales, which could be a hindrance to finding out-of-the-box solutions for intricately complex activities. A direct, straightforward collaboration between TIFR and ISRO might lead to immense friction due to the clash of cultures.

   P C Agrawal's genius of intuitive management style was remarkably efficient in ironing out the above difficulties. He understood the style of the functioning of ISRO. He assiduously cultivated all senior functionaries like P S Goel, the then Director of ISAC and George Joseph, one of the early Dreamers of TIFR who had joined ISRO and had become a senior advisor to ISRO. With K Kasturirangan and all senior functionaries fully supportive of AstroSat, the task looked much more attainable. The authority of these seniors needs to be adequately articulated and percolated down to all working-level personnel of ISRO: an AstroSat Progress Monitoring Committee (APMC) was formed to expressly articulate the technical hurdles and minute the necessary remedial measures. P C Agrawal also understood the individualistic nature of TIFR scientists. He impressed upon the senior scientists of TIFR that AstroSat is a national project and that TIFR should fully support and commit to it. He assigned separate and individual responsibilities to independent scientists like KP and me. He impressed upon the IR group to participate in AstroSat: they did some work for the UVIT.

UVIT is one of the finest examples of technical excellence of the highest order. S N Tandon is the prime force behind UVIT.

UV radiation is the shortest wavelength at which conventional mirrors work (at shorter wavelengths, like in X-rays, one needs grazing incidence mirrors) and being of the shorter wavelength, the mirrors have to be polished to extreme levels of smoothness. ISRO had the expertise to make mirrors to be used in space for remote sensing measurements in optical and IR wavelengths. S N Tandon assiduously worked with the ISRO experts to improve the mirror polish. He established a calibration centre at IIA Bengaluru to test these mirrors. One of the critical features of UVIT is its ability to pinpoint sources in a large field-of-view mirror - an intricate task that has yet to be achieved in UV wavelengths. Every aspect of the satellite system has to be examined and controlled: the pointing of the spacecraft and the jitter associated with it; extreme care for thermal balance; minutely looking at every aspect of the satellite system from the perspective of contributing to the contamination. It was a revelation to see how S N Tandon handled the various ISRO experts: based on first principles, he could understand the validity of the results obtained by their complex modelling and analysis. Though they initially resisted the blunt way S N Tandon opined and corrected the results obtained in their specialist domains, they soon understood the merit of his thinking. Eventually, all intricate aspects were smoothened and worked, and a truly outstanding instrument was built.

\subsection{The making of AstroSat}

After nationwide consultations, a neat configuration was hammered out for AstroSat. Three identical Large Area X-ray Proportional Counters (LAXPC) with the largest effective area ever flown in a wide energy range, along with the Ultra- violet Imaging Telescope (UVIT), would form the heart of the satellite. Complementary X-ray data at low energies would be provided by SXT and at high energies by CZT Imager (CZTI), thus providing precious wide-band spectroscopic data for bright X-ray sources. All these four instruments are co-aligned, and auxiliary data would be provided by a Scanning Sky Monitor (SSM) mounted in an orthogonal direction to scan the sky and record the onset of transient X- ray sources. The satellite would be placed in a near-equatorial orbit to achieve low background for X-ray detectors. A large solid-state onboard memory would provide time-tagged information for each photon for all the instruments. Each of the instruments is an incremental advance to similar instruments flown earlier elsewhere, but jointly, they are a formidable ensemble. Any exciting new event in the X-ray sky, like X-ray transients or X-ray outbursts in an X-ray source, will be noticed by SSM and announced to the world. Exciting events among them would be, later on, continuously stared at by the other instruments to understand and elicit their mysteries. SXT could, unlike earlier focussing X-ray detectors, see very bright sources without distorting the information due to a process called pile-up, thus enabling good wide-band spectral measurements of bright X-ray transients. The extremely wide-viewed UVIT, with its superior seeing compared to its predecessors like the Galex mission, would be very useful in studying extended objects far into the universe. CZTI provided hard X-ray (above $\sim$80 keV) monitoring and hard X-ray polarisation - a new feature with exciting scientific possibilities.

AstroSat was given in-principle approval in 2000 and formally approved in 2003 with a mandate of completing the project and launching it in 2006.

\begin{figure}[h]
\centering
\includegraphics[angle=-90, width=\textwidth]{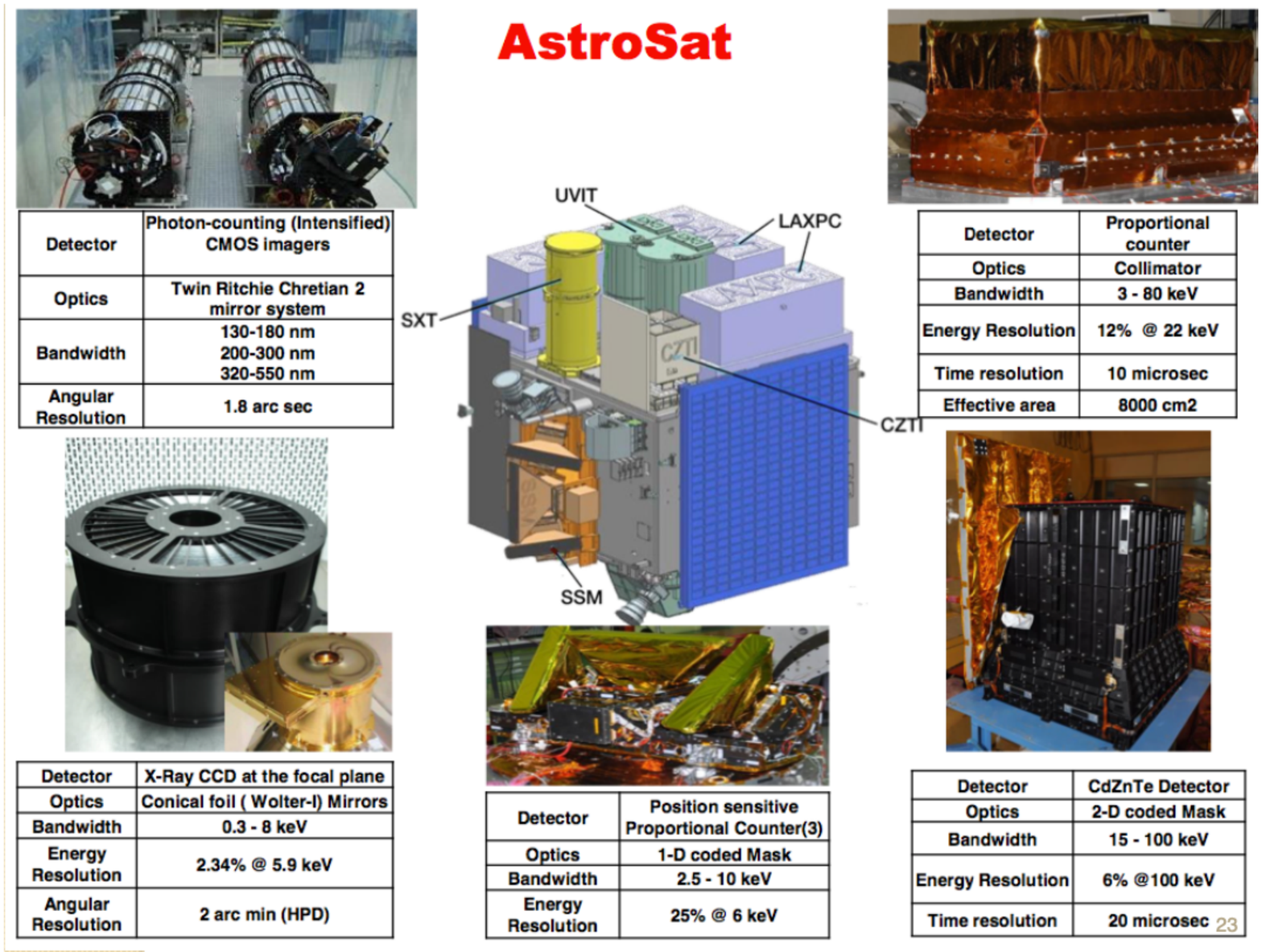} 
\caption{The configuration of AstroSat.}
\label{fig5}
\end{figure}

It took fifteen years to complete the project.

Making proportional counters is a complex art. LAXPC was several orders of magnitude more complex than PPC onboard IXAE: ten times more massive, several times more wires and physically huge and unwieldy. The sheer physical labour is vast: stitching together the delicate wires, manufacturing the pressure vessels from single large blocks of aluminium, making them leakproof, and test- ing for long life. Additionally, one complete configuration was demonstrated for appropriate functioning in a balloon flight. It took a long time to fabricate, test, and validate three units of LAXPC detectors. For the SXT, the painstaking efforts required to coat with gold, cure it, test it, and assemble a vast number of thin aluminium foils used for reflecting X-rays through grazing incidence is truly a labour of love for K P Singh. Of course, S N Tandon will not tolerate even an infinitesimal slippage in quality: he would instead redo things than accept an inferior product. Assembling a large number (64) of CZT detectors was a colossal task. A sort of rehearsal was made by flying a small number (3) of CZT detectors in a Russian experiment (launched in 2009). The typical uncertainties of building proportional counters bogged down SSM too.

By 2013 the situation looked bleak.

P C Agrawal had retired and passed on the baton to R K Manchanda (who had tremendous experience building proportional counters), and Manchanda, too, retired. Though all the essential detectors and telescopes were, more or less, in the final stages of assembly, the arduous task of integrating all the subsystems without any mutual interference, ensuring proper alignment, testing the whole satellite under the simulated severe space conditions etc., were highly daunting tasks. All the attention was focused on ensuring the delivery of the instruments. Hence, the auxiliary but extremely essential tasks like developing ground software, training and grooming the user base, making the appropriate web-based analysis and observational tools etc., were given the short shift. Even if AstroSat was launched, it was difficult to envisage a situation where all the instruments worked as per the design (space is very unforgiving, even the slightest mistake can doom the whole project), and data were analysed quickly and meaningfully.

But miracles do happen.

It was an excellent example of all the positive aspects of the Indian science ecosystem for space coming together. Once each instrument's basic detection systems are reasonably ready, all the remaining works are at the ISRO satellite centre (ISAC) in Bengaluru for testing, integration, and validation. The ISRO hierarchy decided to bring to fruition this long-dragged project. All scientists involved in the detector development literally camped at ISAC. Any problems or shortcomings were discussed and thrashed out on a daily basis. If the problems required redesigning and fabricating some small parts - say an electronics card - it was done in a matter of days, a job which generally takes several months to complete in the ordinary course of time with all the due procedures of quality control and reviews. Remember, none of the procedures was overlooked or ignored - it is just that all the experts are available under one roof round the clock - everything could be done in the best possible efficient way.

\subsection{AstroSat: launch and operation}

Finally, India's first multi-wavelength astronomy satellite was launched from Sriharikota, India, on 2015 September 28. The satellite was of the IRS (Indian Remote Sensing) class, and the launcher was the time-tested PSLV (C30). In a dream-like episode, the satellite was precisely put into a near-circular orbit at an altitude of 650 km and an inclination of 6 degrees. The instrument mass (780 kg) was a good fraction of the satellite mass (1550 kg). The large onboard memory of 200 Gb (transmitted at the rate of 210 Mb/sec) provided unprecedented time- tagged data for each photon for all the instruments. The Satellite Positioning System (SPS) enabled precise time calibration. The satellite had good pointing accuracy of 0.05 degrees and an operational life of $>$ 5 years (still operating after seven years).

Within a month, all the instruments were switched on and made operational. All instruments worked as per the design.

AstroSat is a prime example of productive all-India participation. Most of the ISRO Centres participated, contributing to the satellite, rocket, launch operations, satellite orbit control and management, data analysis software development and data management and providing the overall management of the whole AstroSat project. Further, the test and evaluation of all payloads were carried out at ISRO Satellite Center (ISAC, now URSC), and the SSM was designed, fabricated and delivered at ISAC. CZTI was assembled and tested at VSSC (VSSC also contributed to the CZTI electronics). Research Institutes across India participated in AstroSat: TIFR took the prime responsibility of delivering three instruments (LAXPC, SXT, and CZTI). Indian Institute of Astrophysics (IIA), Bengaluru, was responsible for UVIT. IUCAA, Pune, collaborated in developing SSM and CZTI and took an active interest in the post-launch operation and data analysis. RRI, Bengaluru, collaborated in the development of LAXPC. PRL, Ahmedabad participated in the development of CZTI, particularly the X-ray polarisation aspect of the instrument. Many universities too actively participated in AstroSat activities. Apart from the Indian participation, AstroSat also saw significant contributions from Leicester University (UK) and the Canadian Space Agency.

AstroSat proved its mettle during its seven years of operation. Several ex- citing results were discussed in a meeting held recently at ISRO to celebrate the seven years of AstroSat operations. AstroSat proved to be a versatile multi- wavelength observatory, and the observations resulted in several path-breaking results. The UVIT observations solved a decade-old puzzle of the very hot red star and measured extended emissions from the butterfly nebula. The discovery of the Lyman continuum from a galaxy at a redshift of 1.42 demonstrated the fine imaging quality of UVIT. The celebrated black hole source GRS 1915+105 was observed several times using LAXPC, and a new intermediate state was dis- covered. CZTI measured the off-pulse X-ray polarisation from the Crab pulsar. CZTI also proved an excellent all-sky monitor in hard X-rays recording several GRBs. SXT measured coronal emissions in stars, and the joint LAXPC-SXT measurements recorded several thermo-nuclear bursts in neutron star binaries. Such joint measurements also helped measure the spin of black holes and provided crucial data for multi-wavelength studies of blazars.
   
 \subsection{AstroSat: A personal introspection}
 
AstroSat is India's first observatory-class satellite. Several good practices are generally followed in operating a large observatory class satellite. Since space data are costly to procure, it is necessary to involve the whole scientific community to delve deeply into the data and make the best use of these precious data. Over the past several decades, these methods have been streamlined and made entirely professional for any international space observatory. The standard practices include: 1) a call for proposals to observe specific objects using the observatory and to evaluate these proposals based on a peer-review system so that the best ideas are facilitated with data 2) a careful method to optimally use the satellite time by considering the multitude of constraints pertaining to the technical complexities of manoeuvring the satellite as well as the user requirements based on the peculiarities of the objects that are viewed 3) quickly assessing the quality of the downloaded data and taking corrective actions for any possible limitations in the received data 4) providing the data to the users 5) continuously upgrading the data analysis software for the optimal use of data 6) providing training to new users in data analysis and interpretation and, most importantly, 7) using the data to extract the most meaningful science results from the observations.

All the above-listed tasks might be found to be obviously needed for any observatory class satellite like AstroSat, but being the first from India, getting all these done in time was challenging and arduous work (some of them are still ongoing). Needless to say, the science base in India is small, and some handful of scientists worked in a dedicated manner to bring all these tasks to fruition. Quite often, one can see these A-to-Z scientists working on some software codes in the morning, attending meetings at noon for operating the satellite, organising a Scientific Workshop in the afternoon, and pondering over some science problems arising from looking at the AstroSat data late in the night! It is indeed a great credit to the untiring efforts of these small number of AstroSat scientists that, even after seven years, most of the instruments are working, and the observatory is running smoothly and attracts proposals from scientists all across the globe.

But I have this slight bitter aftertaste: did we miss the woods for the trees?

By the way, in the eighties, when K P Singh and I were doing a joint study on X-ray astronomy, the enormity of the gulf between our efforts and the best results pouring in from international observatories really hit us hard, and each of us resolved to do something. KP resolved to bring X-ray focussing techniques to India and did it for AstroSat. What did I resolve to do? Well, when some major space astronomy instruments are launched, the results are generally published in one issue of The Astrophysical Journal Letters (ApJL), thus making a huge sudden impact on some areas of astrophysics. It was always my dream to be a part of a similar endeavour from India.
  
   AstroSat results did come in single issues of Current Science (immediately after launch - mainly containing instrument description papers) and recently in the Journal of Astronomy and Astrophysics (commemorating the five years of operation of AstroSat). The results, though significant, lack the punch of a collector's issue of ApJL. There are a few epoch-making results from AstroSat, but they are few and far apart.

As far as the instrumental capabilities go, AstroSat does have some niche observational features, which in principle, could have been pushed to the level of making a substantial impact on some selected areas of Astrophysics. This requires a significantly higher level of application: ease of data acquisition, sophistication in analysis software, impactful discussions and a more concentrated scientific approach. One can take a charitable view that the concerned AstroSat scientists were busy delivering the routine tasks of operating AstroSat, leaving little room for a more concentrated approach to scientific investigations. Per- haps. Nevertheless, I will give my personal take on this in the following.

A massive project like AstroSat has magnified the good and bad features of the `TIFR culture'. TIFR encourages individual brilliance and independent thoughts. This is very important to grow expertise in different areas. I would say that AstroSat instruments were built due to the dedicated work of many such individuals: three of the five instruments of AstroSat were led by TIFR scientists; an illustrious alumnus of TIFR led the fourth one (UVIT); the fifth instrument (SSM) had its legacy from a close collaboration between TIFR and ISRO. This individualist capacity simultaneously can prove to be a bane for group activities. It is very rare to see in TIFR  two or more scientists meaningfully collaborating in a sustained manner! No wonder getting more profound scientific results from AstroSat, which needs sizeable collective activity not only to bring together diverse capabilities but also to optimise resource allocation and reduce duplication of work, remains an unrealised dream.

\section{Conclusions}

This article is based on a talk at the TIFR Alumni Association event `Land- marks@TIFR'. This event was organised to celebrate 75 years of independence of India and mainly focused on the milestones TIFR has achieved. AstroSat is one of the significant landmark activities of TIFR. While giving an account of the making of AstroSat, I have put a positive spin on the events to provide a `feel-good' effect to the audience consisting mainly of TIFR alumni. On re-reading the article, I feel that the material covered can also be used to examine the perils and travesties of trying to do first-rate science in a developing country. Perhaps, using the same material, I might give an alternative, more in-depth perspective. Like the classic 1950 movie Rashomon directed by Akira Kurosawa, the same set of events can be constructed into a totally different narrative based on the narrator's perspective!

Another caveat about this article: I have deliberately avoided giving ex- tensive references. Any rigorous referencing would have made the article unwieldily and unreadable. Most of the materials covered here can be obtained from popular reading materials. The readers can find historical accounts of X- ray astronomy in the classic textbooks (\enquote{The X-Ray Universe} by W. Tucker and R. Giacconi; \enquote{X-Ray Astronomy} Edited by R. Giacconi \& H. Gursky). An excellent historical account of the making of AstroSat is given by Hari Pulakkat (\enquote{Space Life Matter} - Hachette Book Publishing India, Pvt Ltd). P C Agrawal's scientific reminiscences  \enquote{Journey through the birth, growth and maturity of X-Ray Astronomy} (RAA, 19, 238, 2019) gives a good historical account of X-ray astronomy at TIFR. However, the rest of the material is based on my personal experiences: it is tough to do rigorous referencing for it.

\end{document}